\documentclass[aps,prl,reprint,superscriptaddress,showpacs,longbibliography,floatfix,footnoteinbib]{revtex4-1}
\usepackage{amsmath,amssymb,graphicx,units}
\usepackage[plainpages=false,pdfpagelabels,colorlinks=true,linkcolor=red,urlcolor=blue,citecolor=blue,pdftitle={Title},pdfauthor={},pdfdisplaydoctitle=true,pdfduplex=DuplexFlipLongEdge]{hyperref}
\usepackage{multirow}
\usepackage{setspace}

\begin{document}
\title{Singlet pairing and superconductivity in $t$-$J$ ladders with Mott insulating stripes}
\author{Chen Cheng}
\affiliation{Beijing Computational Science Research Center, Beijing 100193, China}
\affiliation{Department of Physics, The Pennsylvania State University, University Park, PA 16802, USA}
\author{Rubem Mondaini}
\affiliation{Beijing Computational Science Research Center, Beijing 100193, China}
\author{Marcos Rigol}
\affiliation{Department of Physics, The Pennsylvania State University, University Park, PA 16802, USA}

\begin{abstract}
Pairing in superconductors occurs in a variety of channels and can be produced by various mechanisms. Here, we show that, in the presence of strong correlations, a novel singlet-pair superconducting phase can occur in ladder geometries with the fermions in the pairs residing on different sides of a Mott insulating stripe. The antiferromagnet correlations in the Mott stripe provide the pairing ``glue'' in such a phase. We study, using the density matrix renormalization group method, the ground state of four-leg $t$-$J$ ladders with Mott insulating stripes in the inner two legs. Pairing and superconductivity are revealed by the presence of negative binding energies and by algebraically decaying interleg singlet-pair correlations, respectively.
\end{abstract}

\maketitle

\paragraph{Introduction.}
In spite of being one of the most studied phenomena in condensed matter physics over the past three decades, consensus has not yet been reached as to which is(are) the underlying mechanism(s) for high-temperature superconductivity~\cite{Dagotto1994,Keimer2015}. In high-temperature superconductors, pairing occurs between electrons in the presence of strongly repulsive interactions. A complex competition of different orders makes it difficult to separate which ones aid and which ones are inimical to superconductivity~\cite{Fradkin2015}. One of the orders that has been observed in a variety of experiments in the doped cuprates family is related to the formation of charge density waves (CDW's), dubbed stripes~\cite{Tranquada1995, Tranquada1997, Tranquada2013}.

The presence of strong interactions makes it difficult to solve even the simplified effective models, e.g., the Hubbard and $t$-$J$ models, that have been argued to contain the essential ingredients needed to describe high-temperature superconductivity~\cite{Dagotto1994, Zhang1988, Ogata2008}. Within the last 15 years, a new way to explore the physical phenomena described by those models has emerged in the field of ultracold gases in optical lattices~\cite{BlochRMP2008, esslinger_review_10}, in which artificial lattices are created using laser beams and are loaded with ultracold atoms. This allows experimentalists to engineer nearly ideal realizations of effective model Hamiltonians with remarkable control and tunability, and to potentially identify the phases that can be described by those Hamiltonians. Recent experiments with ultracold fermions in two-dimensional lattices have made great progress in the exploration of the phases described by the two-dimensional Hubbard model at and away from half filling~\cite{Greif2016, Cocchi2016, CheukPRL2016, Parsons2016, Cheuk2016, Drewes2016, Drewes2017, Mazurenko2016, Cocchi2017, Brown2017}, for example.

In experiments with ultracold gases, inhomogeneous trapping potentials (usually generated by the same laser beams that create the artificial lattice) maintain the gas confined. This results in nonuniform density distributions, with the coexistence of space-separated metallic and Mott insulating domains~\cite{rigol_muramatsu_03, rigol_muramatsu_04a, chiesa_varney_11}. While inhomogeneities are generally regarded as a nuance, because one usually would like to understand phases of translationally invariant models, in this Rapid Communication we are interested in properties that are unique to inhomogeneous systems. They could be of relevance to phenomena such as high-temperature superconductivity because of, e.g., the presence of stripes. More specifically, we are interested in the properties of the conducting regions that surround Mott insulating stripes in ladder geometries. Since Mott insulators exhibit antiferromagnetic correlations (in bipartite lattices), negative binding energies can occur \cite{Machida2004, Rigol2005, Machida2005}, and novel superconducting phases can develop~\cite{Emery1997, tsvelik2016a, tsvelik2016b}, because of proximity effects to an antiferromagnet.

\begin{figure}[b] 
 \includegraphics[width=0.99\columnwidth]{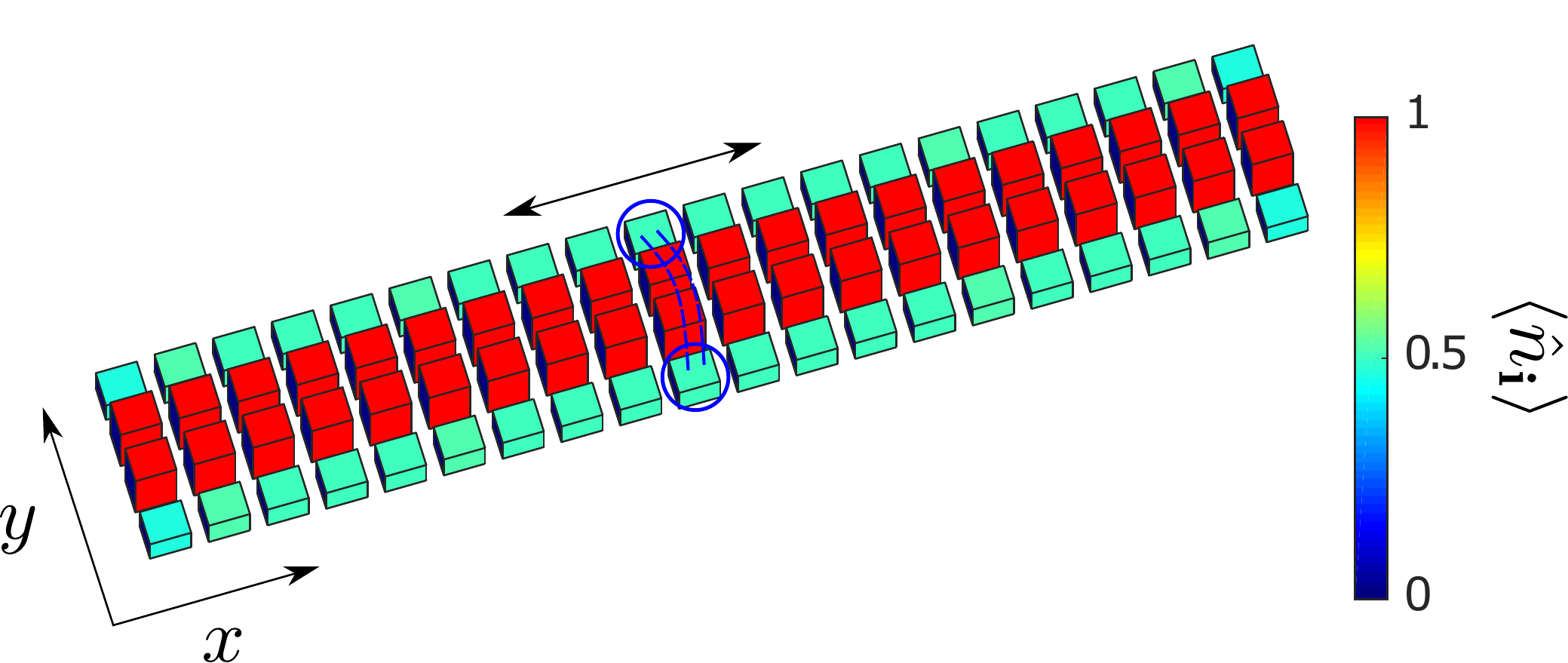}
 \caption{Numerically obtained site-occupation profile, depicted as color bars, in a four-leg ladder with $L_x=20$, $N_\uparrow=N_\downarrow=30$, $V=-40$, for isotropic couplings $J_x=J_y=0.33$ [see Eq.~\eqref{eq_ham1}]. A robust Mott stripe ($\langle \hat n_i \rangle \approx 1$) is present in the inner two legs, while the two outer legs exhibit an average site occupation $\langle \hat n_i \rangle \approx 0.5$. The blue circles, connected by the dashed lines, show the interleg pairing investigated. The arrows indicate the pairs' motion in the superconducting state.}
 \label{fig:fig1}
\end{figure}

Thus, we study the $t$-$J$ model on ladders in which the legs have different on-site potentials. A large negative on-site potential in the inner legs allows us to create Mott insulating stripes with controllable widths. In such systems, we show that anisotropic exchange couplings produce a novel form of pairing in which the paired fermions reside on sites across the Mott insulating stripe (see Fig.~\ref{fig:fig1}). To shed light on the occurrence of pairing and superconductivity, we probe binding energies as well as interleg pairing correlations, for both the singlet and triplet channels. We also study intraleg pairing, one-particle, and spin-spin correlations. The interleg singlet-pair correlations are found to decay algebraically, while the others decay exponentially. These results are contrasted with those obtained in homogeneous two-leg ladders, which exhibit a singlet-pair superconducting phase at low hole doping~\cite{Hayward1995, Poilblanc2003}.

\paragraph{Model and method.}
\label{sec_m_and_m}
The $t$-$J$ Hamiltonian has been extensively studied as a paradigmatic model to understand the effects of strong correlations for large values of the ratio between the on-site repulsion strength $U$ and the hopping parameter $t$~\cite{Zhang1988, Dagotto1994, Hayward1995, White1997, White1998, White2003, Poilblanc2003, Chang2007, rigol_bryant_07b, Ogata2008, rigol_shastry_09b, Moreno2011, Corboz2011}. It can be simulated in optical lattice experiments by trapping two species fermions in deep lattices so that $U/t\gg1$, or using ultracold polar molecules~\cite{Gorshkov2011, Gorshkov2011a}. The $t$-$J$ Hamiltonian reads $\hat H_{tJ} = -t\sum_{\langle {\bf i},{\bf j} \rangle,\sigma} (\hat c^\dagger_{{\bf i},\sigma}\hat c^{}_{{\bf j},\sigma} + {\rm H.c.}) + J\sum_{\langle {\bf i},{\bf j} \rangle} (\vec{S}_{{\bf i}}\cdot\vec{S}_{{\bf j}} - \frac{1}{4}\hat n_{\bf i} \hat n_{\bf j})$, where the operator $\hat c^\dagger_{{\bf i},\sigma}$ ($\hat c_{{\bf i},\sigma}$) creates (annihilates) a fermion with spin $\sigma = \uparrow, \downarrow$ on site ${\bf i}$, and $\langle{\bf i},{\bf j}\rangle$ denotes the constrained summation over pairs of nearest neighbor sites. $\vec{S}_{{\bf i}} = \hat c^\dagger_{{\bf i},s} \vec{\sigma}_{s,s'} \hat c^{}_{{\bf i},s'}$
and $\hat n_{{\bf i}} = \sum_\sigma \hat c^\dagger_{{\bf i},\sigma} \hat c^{}_{{\bf i},\sigma}$ are the spin ($\vec{\sigma}$ are the Pauli matrices) and site occupation operators at site ${\bf i}$, respectively. Note that the fermionic degrees of freedom must be projected onto the Hilbert subspace without double occupancies.

We study the ground state of the $t$-$J$ model in a ladder geometry, with $L_x$ sites in the $x$ direction and $L_y$ sites in the $y$ direction (see Fig.~\ref{fig:fig1}), for which the Hamiltonian can be written as
\begin{eqnarray}
\label{eq_ham1}
\hat {\cal H} &=&
-t_x\sum_{\sigma}\sum_{i_x=1}^{L_x-1}\sum_{i_y=1}^{L_y}\left( \hat c^\dagger_{i_x,i_y,\sigma} \hat c_{i_x+1,i_y,\sigma} + {\rm H.c.}\right)\nonumber \\
&&-t_y\sum_{\sigma}\sum_{i_x=1}^{L_x}\sum_{i_y=1}^{L_y-1}\left( \hat c^\dagger_{i_x,i_y,\sigma} \hat c_{i_x,i_y+1,\sigma} + {\rm H.c.}\right)  \nonumber \\
&&+J_x\sum_{i_x=1}^{L_x-1}\sum_{i_y=1}^{L_y}\left(\vec{S}_{i_x,i_y}\cdot\vec{S}_{i_x+1,i_y} - \frac{1}{4}\hat n_{i_x,i_y}\hat n_{i_x+1,i_y} \right)\nonumber \\
&&+J_y\sum_{i_x=1}^{L_x}\sum_{i_y=1}^{L_y-1}\left(\vec{S}_{i_x,i_y}\cdot\vec{S}_{i_x,i_y+1} - \frac{1}{4}\hat n_{i_x,i_y}\hat n_{i_x,i_y+1} \right) \nonumber\\
&&+\sum_{\bf i} V_{i_y} \hat n_{\bf i}.
\end{eqnarray}
In the remainder of this Rapid Communication, $t_x=t_y=1$ sets the energy scale. We focus on four-leg ladders ($L_y=4$) with an inhomogeneous potential in the $y$ direction. This allows us to create Mott insulating stripes in the inner legs (for which we take $V_{i_y=2}=V_{i_y=3}=V=-40$) while maintaining the filling below one in the outer legs (for which we take $V_{i_y=1}=V_{i_y=4}=0$). 

Note that the outer two legs are not connected by hoppings nor by exchange interactions. A question we address in the following is whether the antiferromagnetic correlations in the inner two Mott insulating legs could induce an effective antiferromagnetic interaction between fermions in the outer two legs resulting in interleg singlet pairing. We consider both isotropic ($J_y=J_x$) and anisotropic ($J_y>J_x$) exchange couplings. In optical lattice experiments, different inter- and intraleg exchange couplings can be engineered by having different inter- and intraleg hopping amplitudes, or, in systems with ultracold polar molecules, by changing the direction of an applied electric field~\cite{Gorshkov2011a}.

The ground state of Eq.~\eqref{eq_ham1} is obtained numerically via the density matrix renormalization group (DMRG) method~\cite{White1992,White1993}. We dynamically use up to $8000$ DMRG many-body states so that the truncation error is of the order of $10^{-7}$~\cite{Legeza2003}. The calculations are done for $J_x=0.33$, a value commonly used in studies of the $t$-$J$ model in two-dimensional lattices~\cite{White1997, White1998, rigol_bryant_07b}, and different values of $J_y$. We focus on four-leg ladders with $N_\uparrow=N_\downarrow$ and average site occupation $(N_\uparrow+N_\downarrow)/(L_xL_y) = 0.75$, so that $\langle \hat n_i \rangle \approx 1$ in the inner two legs and $\langle \hat n_i \rangle \approx 0.5$ in the outer two legs. $N_\uparrow$ ($N_\downarrow$) stands for the number of fermions with spin up (spin down). For computational convenience, open boundary conditions are adopted in the $x$ direction.

\paragraph{Binding energy.}
We start the exploration of the pairing tendencies of the four-leg ladder geometry by examining the binding energy, $E_b = E_0(N_\uparrow+1,N_\downarrow+1) + E_0(N_\uparrow,N_\downarrow) - 2E_0(N_\uparrow+1,N_\downarrow)$, where $E_0(N_\uparrow,N_\downarrow)$ stands for the ground state energy in a system with $N_\uparrow$ ($N_\downarrow$) spin-up (spin-down) fermions. $E_b<0$ in the thermodynamic limit means that the energy of two interacting particles (or holes, depending on the filling) is lower than that of two noninteracting ones. As a result, the system exhibits a tendency toward pair formation.

\begin{figure}[!t] 
 \includegraphics[width=0.99\columnwidth]{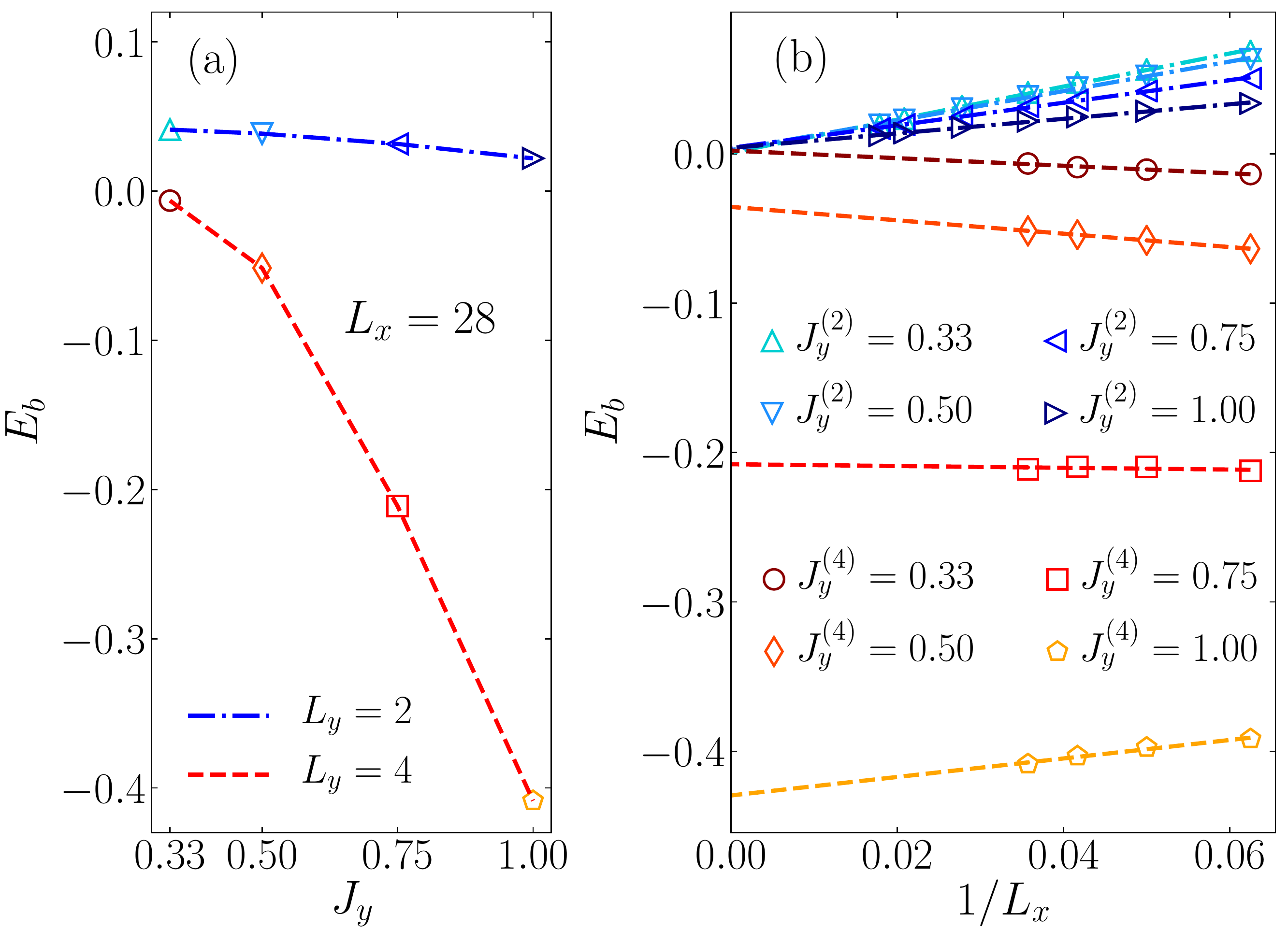}
 \caption{Binding energies $E_b$ in four-leg ladders with a Mott insulating stripe in the inner two legs, and in homogeneous two-leg ladders with the same average site occupation as in the outer legs of the four-leg ladders (see text). (a) $E_b$ as a function of $J_y$ for $L_x=28$. (b) Finite-size scaling analyses of $E_b$; dashed lines are linear fits in $1/L_x$. $J_y^{(L_y)}$ denotes $J_y$ for the system with $L_y$ legs.}
 \label{fig:fig2}
\end{figure}

Figure~\ref{fig:fig2}(a) shows results for $E_b$ vs $J_y$ in four-leg ladders with $L_x=28$. One can see that the binding energy becomes increasingly negative as $J_y$ increases. Finite-size scaling analyses of $E_b$ vs $1/L_x$ are reported in Fig.~\ref{fig:fig2}(b). While for the isotropic ($J_y=J_x=0.33$) case the binding energy appears to be positive (but very small) in the thermodynamic limit, for the anisotropic ones the extrapolations indicate that they exhibit large negative values.

At this point one could ask whether the Mott stripe in between the outer two legs with $\langle \hat n_i \rangle \approx 0.5$ is necessary to observe robust negative binding energies. This is a valid concern as connecting the two outer legs after removing the Mott stripe could lead to similar (or even larger) negative binding energies. Homogeneous two-leg ladders at lower hole doping $\langle \hat n_i \rangle \approx 0.8$ have been shown to support singlet-pair superconductivity~\cite{Hayward1995, Poilblanc2003}. For $L_x=28$ [Fig.~\ref{fig:fig2}(a)], $E_b$ for two-leg ladders can be seen to depend weakly (compared to the results for the four-leg ladder) on the value of $J_y$. Finite-size scaling analyses of $E_b$ for two-leg ladders [Fig.~\ref{fig:fig2}(b)], for the four values of $J_y$ studied, suggest that the binding energies are very small and positive, or vanish, in the thermodynamic limit. 

\paragraph{Pairing correlation functions.}
As mentioned before, a unique feature of our setup is that the Mott stripe in the inner two legs of the four-leg ladder can potentially mediate pairing between fermions on the opposite outer legs. To identify the pairing tendency that is dominant in our system, we compute the interleg and intraleg singlet- and triplet-pair correlation function
\begin{equation}
P_{x_1,x_2} = \langle \hat \Delta^\dagger_{x_1} \hat \Delta_{x_2}\rangle,
\label{eq_pair}
\end{equation}
where $\Delta^\dagger_{x_1}=\frac{1}{\sqrt{2}}(\hat c^\dagger_{x_1,1,\downarrow}\hat c^\dagger_{x_1,Ly,\uparrow} - \hat c^\dagger_{x_1,1,\uparrow}\hat c^\dagger_{x_1,Ly,\downarrow})$ for singlet ($P^S$), and $\Delta^\dagger_{x_1}=c^\dagger_{x_1,1,\downarrow}\hat c^\dagger_{x_1,Ly,\downarrow}$ for triplet ($P^T$), interleg pairing, and $\hat \Delta^\dagger_{x_1}=\frac{1}{\sqrt{2}}(\hat c^\dagger_{x_1,1,\downarrow}\hat c^\dagger_{x_1+1,1,\uparrow} - \hat c^\dagger_{x_1,1,\uparrow}\hat c^\dagger_{x_1+1,1,\downarrow})$ for singlet ($P^S_{1D}$), and $\hat \Delta^\dagger_{x_1,y}=c^\dagger_{x_1,y,\downarrow}\hat c^\dagger_{x_1+1,y,\downarrow}$ for triplet ($P^T_{1D}$), intraleg pairing. (We checked that the results for the triplet-pair correlations are identical if one considers spin-up fermions, and that intraleg correlations are identical in the two outer legs.) Since we use open boundary conditions in our numerical calculations, in what follows we report the average over all correlations at the same distance
\begin{eqnarray}
P(r) = \frac{1}{\cal N}\sum_{|x_1-x_2|=r} P_{x_1,x_2},
\label{eq:ave_corr}
\end{eqnarray}
where ${\cal N}$ is the total number of pairs of sites $\{x_1,x_2\}$ satisfying $|x_1-x_2|=r$.

\begin{figure}[!t] 
 \includegraphics[width=0.99\columnwidth]{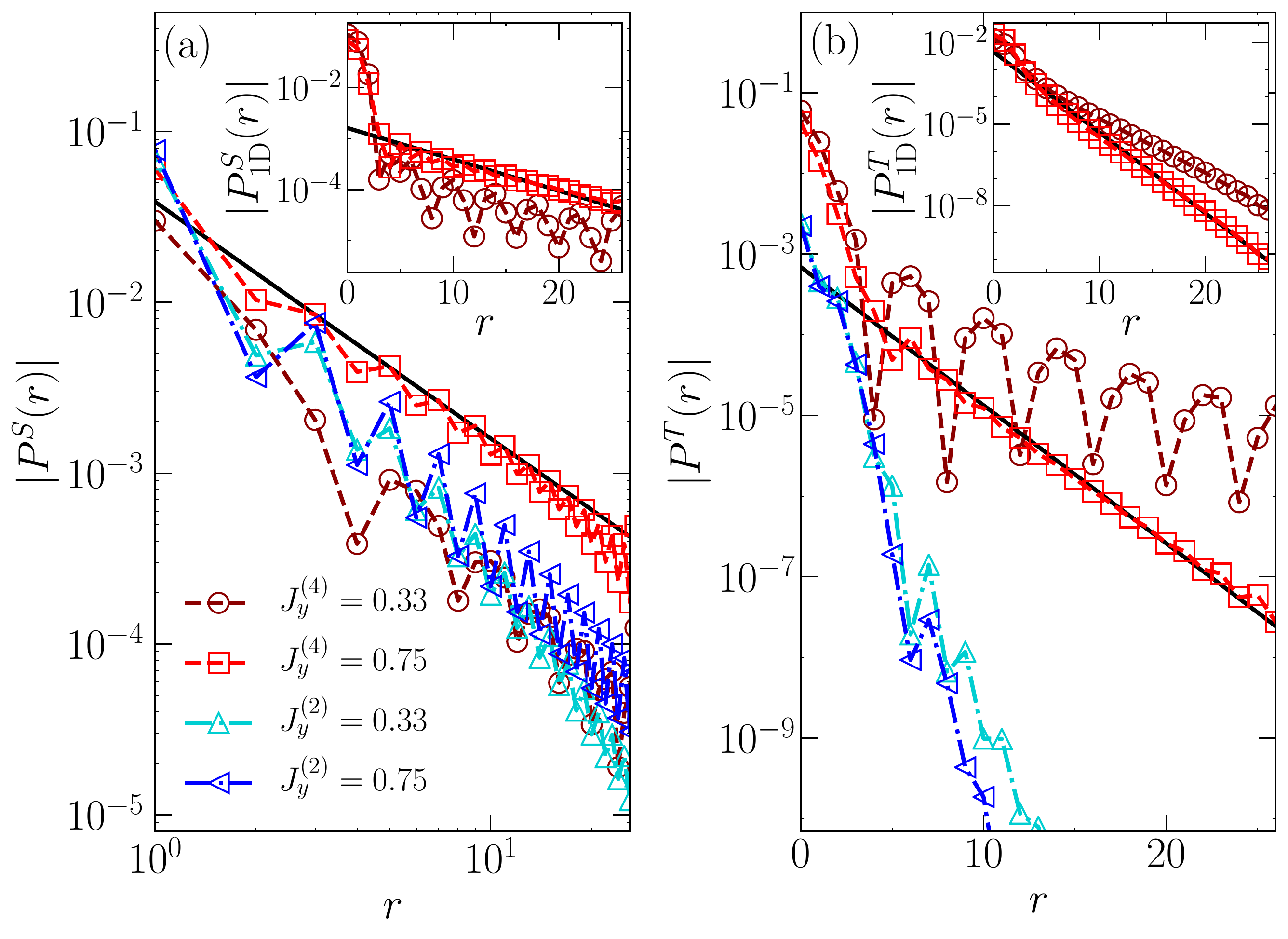}
 \caption{(a) Singlet- and (b) triplet-pair correlation functions in four-leg ladders with a Mott insulating stripe in the inner two legs, and in homogeneous two-leg ladders, with $L_x=28$. The main panels show results for interleg pairing correlations, while the insets show results for intraleg pairing correlations in the four-leg ladders. Note the log-log scale in the main panel in (a) vs the log-linear scale in all other plots. The black solid lines depict results of power-law [main panel in (a)] and exponential [main panel in (b) and both insets] fits.}
 \label{fig:fig3}
\end{figure}

In Fig.~\ref{fig:fig3}(a) we plot the interleg singlet-pair correlations versus $r$ in four-leg ladders. For the isotropic case, their decay is approximately algebraic. Increasing the value of $J_y$ results in an enhancement of those correlations. This is the opposite to what happens for the triplet interleg correlations, depicted in Fig.~\ref{fig:fig3}(b). Their decay is also close to algebraic in the isotropic case, but increasing the value of $J_y$ results in a clear exponential decay. The insets in Figs.~\ref{fig:fig3}(a) and~\ref{fig:fig3}(b) show the intraleg singlet- and triplet-pair correlations for the same values of $J_y$ as in the main panels. They can also be seen to decay exponentially with $r$. These results suggest that, for anisotropic exchange couplings, four-leg ladders with a Mott stripe in the inner two legs exhibit singlet-pair superconductivity with the fermions in the pair being on opposite legs about the Mott insulating stripe.

In Figs.~\ref{fig:fig3}(a) and~\ref{fig:fig3}(b), we also plot the interleg singlet- and triplet-pair correlations, respectively, in homogeneous two-leg ladders. Figure~\ref{fig:fig3}(a) shows that for isotropic exchange couplings the interleg singlet-pair correlations in the two-leg ladder decay slightly faster than those in the four-leg ladder. An anisotropy in the exchange couplings results in a slower decay of the interleg singlet correlations. For $J_y=0.75$ at the largest distance accessible to us, the interleg singlet-pair correlations are nearly an order of magnitude smaller than those in the four-leg ladder. The decay of the interleg triplet correlations [Fig.~\ref{fig:fig3}(b)], as well as of the intraleg singlet and triplet-pair correlations (not shown), is exponential in two-leg ladders.

\paragraph{One-particle and spin-spin correlations.}
The quantum many-body phase realized in four-leg ladders with a Mott stripe in the inner legs can be further differentiated from the one in two-leg ladders with $\langle \hat n_i \rangle \approx 0.5$ by studying the intraleg one-particle and spin-spin correlations. In a Luttinger liquid, a relevant point of reference for our systems as we are dealing with quasi-one-dimensional geometries, those correlations exhibit an algebraic decay~\cite{Giamarchibook}. 

We compute the one-particle density matrix
\begin{equation}
\rho_{x_1,x_2} = \langle \hat c_{x_1,1,\downarrow}^\dagger \hat c_{x_2,1,\downarrow} \rangle
\end{equation}
(the results for spin-up fermions are identical), and the spin-spin correlation function
\begin{equation}
S^z_{x_1,x_2} = \langle \hat S^z_{x_1,1} \hat S^z_{x_2,1} \rangle,
\end{equation}
and report averages over correlations at the same distance, which are calculated as in Eq.~\eqref{eq:ave_corr}.

\begin{figure}[!t] 
 \includegraphics[width=0.99\columnwidth]{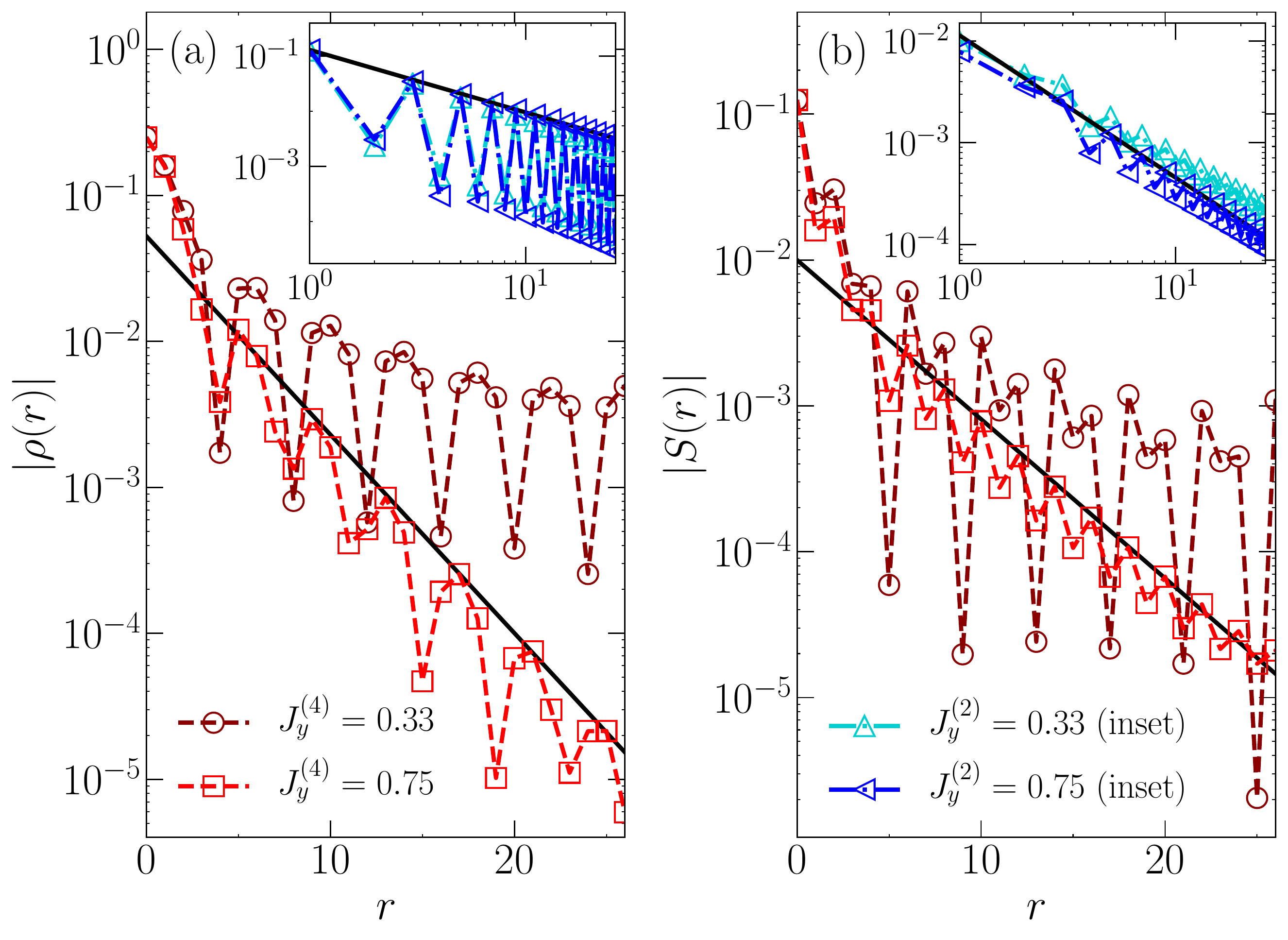}
 \caption{(a) One-particle and (b) spin-spin correlation functions in four-leg ladders with a Mott insulating stripe in the inner two legs (main panels), and in homogeneous two-leg ladders (insets), with $L_x=28$. Note the log-linear scale in the main panels vs the log-log scale in the insets. The black solid lines depict results of exponential (main panels) and power-law (insets) fits.}
 \label{fig:fig4}
\end{figure}

In Figs.~\ref{fig:fig4}(a) and~\ref{fig:fig4}(b), we show results obtained for $\rho(r)$ and $S(r)$, respectively, in four-leg ladders with a Mott stripe (for the same values of $J_y$ as in Fig.~\ref{fig:fig3}). For the isotropic case, both correlation functions exhibit a near algebraic decay with $r$. However, for $J_y=0.75$, $\rho(r)$ and $S(r)$ can be seen to decay exponentially. This is the result of the single-particle and spin excitations being gapped in the singlet-pair superconducting phase.

The insets in Figs.~\ref{fig:fig4}(a) and~\ref{fig:fig4}(b) show results for the same correlation functions in homogeneous two-leg ladders with $\langle \hat n_i \rangle \approx 0.5$. In stark contrast to the results for four-leg ladders with a Mott stripe in the inner legs, and to the results in Ref.~\cite{Hayward1995} in homogeneous two-leg ladders with $\langle \hat n_i \rangle \approx 0.8$, in the two-leg ladders with $\langle \hat n_i \rangle \approx 0.5$ one can see that $\rho(r)$ and $S(r)$ decay algebraically with $r$. Hence, in the anisotropic case in two-leg ladders (at $\langle \hat n_i \rangle \approx 0.5$) there is a competition between one-particle, spin-spin, and interleg singlet-pair correlations, all of which are found to decay algebraically. There is no such competition in four-leg ladders with a Mott stripe in the inner legs in which the one-particle and spin-spin correlations decay exponentially.

The contrast between the binding energies and correlations in two- and four-leg ladders highlights the importance of the Mott stripe in the inner legs of the four-leg ladders for the occurrence of singlet-pair superconductivity between legs with $\langle \hat n_i \rangle \approx 0.5$. Not only does the Mott insulating stripe induce antiferromagnetic interactions between the outer two legs producing interleg singlet-pairing, but, because of the no double-occupancy constraint, the Mott stripe restricts the motion of the fermions in the outer legs to be one dimensional. This stabilizes interleg singlet pairs against delocalization across the legs and makes the interleg singlet-pair superconducting phase robust against doping.

\paragraph{Summary and discussion.}
We have shown that a novel singlet-pair superconducting phase can occur in inhomogeneous $t$-$J$ ladders with Mott insulating stripes in the inner legs. Pairing is mediated by the antiferromagnetic correlations present in the Mott stripe (replacing the Mott insulating stripe by a band insulating one would result in no pairing). What is most remarkable about the superconducting phase discussed here, and contrasts the superconductivity induced by proximity effects to magnetic insulators discussed in other studies~\cite{Emery1997, tsvelik2016a, tsvelik2016b, Rohling2018}, is that the pairs are composed of fermions that are on different sides of the Mott stripe. Hence, the superconducting current involves the two legs on the sides of the Mott stripe. Blocking one of those legs would destroy the supercurrent, an effect that could be used to reveal the existence of the nonlocal pairing across the Mott stripe.

The singlet-pair superconducting phase is characterized by a negative binding energy, algebraically decaying interleg (between the outer legs) singlet-pair correlations; and exponentially decaying interleg triplet-pair correlations, intraleg singlet- and triplet-pair correlations, as well as intraleg one-particle and spin-spin correlations. We also studied (not shown) the intraleg density-density correlations in this phase. As the singlet-pair superconducting phase in homogeneous two-leg ladders at $\langle \hat n_i \rangle \approx 0.8$~\cite{Hayward1995, Poilblanc2003}, in our four-leg ladders the density-density correlations decay algebraically but more rapidly than the interleg singlet-pair correlations.

Since antiferromagnet correlations across Mott insulating stripes are robust against an increase in the number of legs, we expect the singlet-pair superconducting phase found in four-leg ladders with a Mott stripe in the two inner legs to occur in wider ladders with a larger number of even inner legs forming the Mott stripe~\footnote{In ladders with an odd number of legs, we find that there is no novel interleg superconductivity of the type discussed here for ladders with an even number of legs. interleg singlet pairing between fermions in the outer legs is incompatible with antiferromagnetic correlations in an inner Mott insulating stripe with an odd number of legs, and triplet pairing does not appear to be energetically favorable in those configurations.}. This, as well as the effect of changing the anisotropy in the exchange couplings and doping in wider ladders, is something that could be explored using optical lattice experiments.

\begin{acknowledgments}
  M.R. is grateful to A. Muramatsu, R. T. Scalettar, and R. R. P. Singh for motivating discussions, and to S. A. Kivelson, M. M. Maska, and A. Tsvelik for bringing to our attention relevant references. We acknowledge support from NSAF Grant No.~U1530401 (C.C. and R.M.), NSFC Grants No.~11674021 (C.C. and R.M.), and No.~11650110441 (R.M.), and NSF Grant No.~PHY-1707482 (M.R.). The computations were performed in the Tianhe-2JK at the Beijing Computational Science Research Center (CSRC).
\end{acknowledgments}
\bibliography{tj_ladder_refs}

\begin{thebibliography}{47}%
\makeatletter
\providecommand \@ifxundefined [1]{%
 \@ifx{#1\undefined}
}%
\providecommand \@ifnum [1]{%
 \ifnum #1\expandafter \@firstoftwo
 \else \expandafter \@secondoftwo
 \fi
}%
\providecommand \@ifx [1]{%
 \ifx #1\expandafter \@firstoftwo
 \else \expandafter \@secondoftwo
 \fi
}%
\providecommand \natexlab [1]{#1}%
\providecommand \enquote  [1]{``#1''}%
\providecommand \bibnamefont  [1]{#1}%
\providecommand \bibfnamefont [1]{#1}%
\providecommand \citenamefont [1]{#1}%
\providecommand \href@noop [0]{\@secondoftwo}%
\providecommand \href [0]{\begingroup \@sanitize@url \@href}%
\providecommand \@href[1]{\@@startlink{#1}\@@href}%
\providecommand \@@href[1]{\endgroup#1\@@endlink}%
\providecommand \@sanitize@url [0]{\catcode `\\12\catcode `\$12\catcode
  `\&12\catcode `\#12\catcode `\^12\catcode `\_12\catcode `\%12\relax}%
\providecommand \@@startlink[1]{}%
\providecommand \@@endlink[0]{}%
\providecommand \url  [0]{\begingroup\@sanitize@url \@url }%
\providecommand \@url [1]{\endgroup\@href {#1}{\urlprefix }}%
\providecommand \urlprefix  [0]{URL }%
\providecommand \Eprint [0]{\href }%
\providecommand \doibase [0]{http://dx.doi.org/}%
\providecommand \selectlanguage [0]{\@gobble}%
\providecommand \bibinfo  [0]{\@secondoftwo}%
\providecommand \bibfield  [0]{\@secondoftwo}%
\providecommand \translation [1]{[#1]}%
\providecommand \BibitemOpen [0]{}%
\providecommand \bibitemStop [0]{}%
\providecommand \bibitemNoStop [0]{.\EOS\space}%
\providecommand \EOS [0]{\spacefactor3000\relax}%
\providecommand \BibitemShut  [1]{\csname bibitem#1\endcsname}%
\let\auto@bib@innerbib\@empty
\bibitem [{\citenamefont {Dagotto}(1994)}]{Dagotto1994}%
  \BibitemOpen
  \bibfield  {author} {\bibinfo {author} {\bibfnamefont {E.}~\bibnamefont
  {Dagotto}},\ }\bibfield  {title} {\enquote {\bibinfo {title} {Correlated
  electrons in high-temperature superconductors},}\ }\href {\doibase
  10.1103/RevModPhys.66.763} {\bibfield  {journal} {\bibinfo  {journal} {Rev.
  Mod. Phys.}\ }\textbf {\bibinfo {volume} {66}},\ \bibinfo {pages} {763--840}
  (\bibinfo {year} {1994})}\BibitemShut {NoStop}%
\bibitem [{\citenamefont {Keimer}\ \emph {et~al.}(2015)\citenamefont {Keimer},
  \citenamefont {Kivelson}, \citenamefont {Norman}, \citenamefont {Uchida},\
  and\ \citenamefont {Zaanen}}]{Keimer2015}%
  \BibitemOpen
  \bibfield  {author} {\bibinfo {author} {\bibfnamefont {B.}~\bibnamefont
  {Keimer}}, \bibinfo {author} {\bibfnamefont {S.~A.}\ \bibnamefont
  {Kivelson}}, \bibinfo {author} {\bibfnamefont {M.~R.}\ \bibnamefont
  {Norman}}, \bibinfo {author} {\bibfnamefont {S.}~\bibnamefont {Uchida}}, \
  and\ \bibinfo {author} {\bibfnamefont {J.}~\bibnamefont {Zaanen}},\
  }\bibfield  {title} {\enquote {\bibinfo {title} {From quantum matter to
  high-temperature superconductivity in copper oxides},}\ }\href
  {http://dx.doi.org/10.1038/nature14165} {\bibfield  {journal} {\bibinfo
  {journal} {Nature}\ }\textbf {\bibinfo {volume} {518}},\ \bibinfo {pages}
  {179} (\bibinfo {year} {2015})}\BibitemShut {NoStop}%
\bibitem [{\citenamefont {Fradkin}\ \emph {et~al.}(2015)\citenamefont
  {Fradkin}, \citenamefont {Kivelson},\ and\ \citenamefont
  {Tranquada}}]{Fradkin2015}%
  \BibitemOpen
  \bibfield  {author} {\bibinfo {author} {\bibfnamefont {E.}~\bibnamefont
  {Fradkin}}, \bibinfo {author} {\bibfnamefont {S.~A.}\ \bibnamefont
  {Kivelson}}, \ and\ \bibinfo {author} {\bibfnamefont {J.~M.}\ \bibnamefont
  {Tranquada}},\ }\bibfield  {title} {\enquote {\bibinfo {title} {Colloquium:
  Theory of intertwined orders in high temperature superconductors},}\ }\href
  {\doibase 10.1103/RevModPhys.87.457} {\bibfield  {journal} {\bibinfo
  {journal} {Rev. Mod. Phys.}\ }\textbf {\bibinfo {volume} {87}},\ \bibinfo
  {pages} {457--482} (\bibinfo {year} {2015})}\BibitemShut {NoStop}%
\bibitem [{\citenamefont {Tranquada}\ \emph {et~al.}(1995)\citenamefont
  {Tranquada}, \citenamefont {Sternlieb}, \citenamefont {Axe}, \citenamefont
  {Nakamura},\ and\ \citenamefont {Uchida}}]{Tranquada1995}%
  \BibitemOpen
  \bibfield  {author} {\bibinfo {author} {\bibfnamefont {J.~M.}\ \bibnamefont
  {Tranquada}}, \bibinfo {author} {\bibfnamefont {B.~J.}\ \bibnamefont
  {Sternlieb}}, \bibinfo {author} {\bibfnamefont {J.~D.}\ \bibnamefont {Axe}},
  \bibinfo {author} {\bibfnamefont {Y.}~\bibnamefont {Nakamura}}, \ and\
  \bibinfo {author} {\bibfnamefont {S.}~\bibnamefont {Uchida}},\ }\bibfield
  {title} {\enquote {\bibinfo {title} {Evidence for stripe correlations of
  spins and holes in copper oxide superconductors},}\ }\href {\doibase
  10.1038/375561a0} {\bibfield  {journal} {\bibinfo  {journal} {Nature}\
  }\textbf {\bibinfo {volume} {375}},\ \bibinfo {pages} {561} (\bibinfo {year}
  {1995})}\BibitemShut {NoStop}%
\bibitem [{\citenamefont {Tranquada}\ \emph {et~al.}(1997)\citenamefont
  {Tranquada}, \citenamefont {Axe}, \citenamefont {Ichikawa}, \citenamefont
  {Moodenbaugh}, \citenamefont {Nakamura},\ and\ \citenamefont
  {Uchida}}]{Tranquada1997}%
  \BibitemOpen
  \bibfield  {author} {\bibinfo {author} {\bibfnamefont {J.~M.}\ \bibnamefont
  {Tranquada}}, \bibinfo {author} {\bibfnamefont {J.~D.}\ \bibnamefont {Axe}},
  \bibinfo {author} {\bibfnamefont {N.}~\bibnamefont {Ichikawa}}, \bibinfo
  {author} {\bibfnamefont {A.~R.}\ \bibnamefont {Moodenbaugh}}, \bibinfo
  {author} {\bibfnamefont {Y.}~\bibnamefont {Nakamura}}, \ and\ \bibinfo
  {author} {\bibfnamefont {S.}~\bibnamefont {Uchida}},\ }\bibfield  {title}
  {\enquote {\bibinfo {title} {Coexistence of, and competition between,
  superconductivity and charge-stripe order in ${{\rm
  La}}_{1.6\ensuremath{-}\mathit{x}}{{\rm Nd}}_{0.4}{{\rm
  Sr}}_{\mathit{x}}{{\rm CuO}}_{4}$},}\ }\href {\doibase
  10.1103/PhysRevLett.78.338} {\bibfield  {journal} {\bibinfo  {journal} {Phys.
  Rev. Lett.}\ }\textbf {\bibinfo {volume} {78}},\ \bibinfo {pages} {338}
  (\bibinfo {year} {1997})}\BibitemShut {NoStop}%
\bibitem [{\citenamefont {Tranquada}(2013)}]{Tranquada2013}%
  \BibitemOpen
  \bibfield  {author} {\bibinfo {author} {\bibfnamefont {J.~M.}\ \bibnamefont
  {Tranquada}},\ }\bibfield  {title} {\enquote {\bibinfo {title} {Spins,
  stripes, and superconductivity in hole-doped cuprates},}\ }\href {\doibase
  10.1063/1.4818402} {\bibfield  {journal} {\bibinfo  {journal} {AIP Conference
  Proceedings}\ }\textbf {\bibinfo {volume} {1550}},\ \bibinfo {pages}
  {114--187} (\bibinfo {year} {2013})}\BibitemShut {NoStop}%
\bibitem [{\citenamefont {Zhang}\ and\ \citenamefont {Rice}(1988)}]{Zhang1988}%
  \BibitemOpen
  \bibfield  {author} {\bibinfo {author} {\bibfnamefont {F.~C.}\ \bibnamefont
  {Zhang}}\ and\ \bibinfo {author} {\bibfnamefont {T.~M.}\ \bibnamefont
  {Rice}},\ }\bibfield  {title} {\enquote {\bibinfo {title} {Effective
  {H}amiltonian for the superconducting {Cu} oxides},}\ }\href {\doibase
  10.1103/PhysRevB.37.3759} {\bibfield  {journal} {\bibinfo  {journal} {Phys.
  Rev. B}\ }\textbf {\bibinfo {volume} {37}},\ \bibinfo {pages} {3759--3761}
  (\bibinfo {year} {1988})}\BibitemShut {NoStop}%
\bibitem [{\citenamefont {Ogata}\ and\ \citenamefont
  {Fukuyama}(2008)}]{Ogata2008}%
  \BibitemOpen
  \bibfield  {author} {\bibinfo {author} {\bibfnamefont {M.}~\bibnamefont
  {Ogata}}\ and\ \bibinfo {author} {\bibfnamefont {H.}~\bibnamefont
  {Fukuyama}},\ }\bibfield  {title} {\enquote {\bibinfo {title} {The {$t-J$}
  model for the oxide high-{$T_c$} superconductors},}\ }\href
  {http://stacks.iop.org/0034-4885/71/i=3/a=036501} {\bibfield  {journal}
  {\bibinfo  {journal} {Reports on Progress in Physics}\ }\textbf {\bibinfo
  {volume} {71}},\ \bibinfo {pages} {036501} (\bibinfo {year}
  {2008})}\BibitemShut {NoStop}%
\bibitem [{\citenamefont {Bloch}\ \emph {et~al.}(2008)\citenamefont {Bloch},
  \citenamefont {Dalibard},\ and\ \citenamefont {Zwerger}}]{BlochRMP2008}%
  \BibitemOpen
  \bibfield  {author} {\bibinfo {author} {\bibfnamefont {I.}~\bibnamefont
  {Bloch}}, \bibinfo {author} {\bibfnamefont {J.}~\bibnamefont {Dalibard}}, \
  and\ \bibinfo {author} {\bibfnamefont {W.}~\bibnamefont {Zwerger}},\
  }\bibfield  {title} {\enquote {\bibinfo {title} {Many-body physics with
  ultracold gases},}\ }\href {\doibase 10.1103/RevModPhys.80.885} {\bibfield
  {journal} {\bibinfo  {journal} {Rev. Mod. Phys.}\ }\textbf {\bibinfo {volume}
  {80}},\ \bibinfo {pages} {885--964} (\bibinfo {year} {2008})}\BibitemShut
  {NoStop}%
\bibitem [{\citenamefont {Esslinger}(2010)}]{esslinger_review_10}%
  \BibitemOpen
  \bibfield  {author} {\bibinfo {author} {\bibfnamefont {T.}~\bibnamefont
  {Esslinger}},\ }\bibfield  {title} {\enquote {\bibinfo {title}
  {{Fermi-Hubbard} physics with atoms in an optical lattice},}\ }\href@noop {}
  {\bibfield  {journal} {\bibinfo  {journal} {Annual Review of Condensed Matter
  Physics}\ }\textbf {\bibinfo {volume} {1}},\ \bibinfo {pages} {129--152}
  (\bibinfo {year} {2010})}\BibitemShut {NoStop}%
\bibitem [{\citenamefont {Greif}\ \emph {et~al.}(2016)\citenamefont {Greif},
  \citenamefont {Parsons}, \citenamefont {Mazurenko}, \citenamefont {Chiu},
  \citenamefont {Blatt}, \citenamefont {Huber}, \citenamefont {Ji},\ and\
  \citenamefont {Greiner}}]{Greif2016}%
  \BibitemOpen
  \bibfield  {author} {\bibinfo {author} {\bibfnamefont {D.}~\bibnamefont
  {Greif}}, \bibinfo {author} {\bibfnamefont {M.~F.}\ \bibnamefont {Parsons}},
  \bibinfo {author} {\bibfnamefont {A.}~\bibnamefont {Mazurenko}}, \bibinfo
  {author} {\bibfnamefont {C.~S.}\ \bibnamefont {Chiu}}, \bibinfo {author}
  {\bibfnamefont {S.}~\bibnamefont {Blatt}}, \bibinfo {author} {\bibfnamefont
  {F.}~\bibnamefont {Huber}}, \bibinfo {author} {\bibfnamefont
  {G.}~\bibnamefont {Ji}}, \ and\ \bibinfo {author} {\bibfnamefont
  {M.}~\bibnamefont {Greiner}},\ }\bibfield  {title} {\enquote {\bibinfo
  {title} {Site-resolved imaging of a fermionic {M}ott insulator},}\ }\href
  {\doibase 10.1126/science.aad9041} {\bibfield  {journal} {\bibinfo  {journal}
  {Science}\ }\textbf {\bibinfo {volume} {351}},\ \bibinfo {pages} {953--957}
  (\bibinfo {year} {2016})}\BibitemShut {NoStop}%
\bibitem [{\citenamefont {Cocchi}\ \emph {et~al.}(2016)\citenamefont {Cocchi},
  \citenamefont {Miller}, \citenamefont {Drewes}, \citenamefont {Koschorreck},
  \citenamefont {Pertot}, \citenamefont {Brennecke},\ and\ \citenamefont
  {K\"ohl}}]{Cocchi2016}%
  \BibitemOpen
  \bibfield  {author} {\bibinfo {author} {\bibfnamefont {E.}~\bibnamefont
  {Cocchi}}, \bibinfo {author} {\bibfnamefont {L.~A.}\ \bibnamefont {Miller}},
  \bibinfo {author} {\bibfnamefont {J.~H.}\ \bibnamefont {Drewes}}, \bibinfo
  {author} {\bibfnamefont {M.}~\bibnamefont {Koschorreck}}, \bibinfo {author}
  {\bibfnamefont {D.}~\bibnamefont {Pertot}}, \bibinfo {author} {\bibfnamefont
  {F.}~\bibnamefont {Brennecke}}, \ and\ \bibinfo {author} {\bibfnamefont
  {M.}~\bibnamefont {K\"ohl}},\ }\bibfield  {title} {\enquote {\bibinfo {title}
  {Equation of state of the two-dimensional {H}ubbard model},}\ }\href
  {\doibase 10.1103/PhysRevLett.116.175301} {\bibfield  {journal} {\bibinfo
  {journal} {Phys. Rev. Lett.}\ }\textbf {\bibinfo {volume} {116}},\ \bibinfo
  {pages} {175301} (\bibinfo {year} {2016})}\BibitemShut {NoStop}%
\bibitem [{\citenamefont {Cheuk}\ \emph
  {et~al.}(2016{\natexlab{a}})\citenamefont {Cheuk}, \citenamefont {Nichols},
  \citenamefont {Lawrence}, \citenamefont {Okan}, \citenamefont {Zhang},\ and\
  \citenamefont {Zwierlein}}]{CheukPRL2016}%
  \BibitemOpen
  \bibfield  {author} {\bibinfo {author} {\bibfnamefont {L.~W.}\ \bibnamefont
  {Cheuk}}, \bibinfo {author} {\bibfnamefont {M.~A.}\ \bibnamefont {Nichols}},
  \bibinfo {author} {\bibfnamefont {K.~R.}\ \bibnamefont {Lawrence}}, \bibinfo
  {author} {\bibfnamefont {M.}~\bibnamefont {Okan}}, \bibinfo {author}
  {\bibfnamefont {H.}~\bibnamefont {Zhang}}, \ and\ \bibinfo {author}
  {\bibfnamefont {M.~W.}\ \bibnamefont {Zwierlein}},\ }\bibfield  {title}
  {\enquote {\bibinfo {title} {Observation of {2D} fermionic {M}ott insulators
  of $^{40}\mathrm{K}$ with single-site resolution},}\ }\href {\doibase
  10.1103/PhysRevLett.116.235301} {\bibfield  {journal} {\bibinfo  {journal}
  {Phys. Rev. Lett.}\ }\textbf {\bibinfo {volume} {116}},\ \bibinfo {pages}
  {235301} (\bibinfo {year} {2016}{\natexlab{a}})}\BibitemShut {NoStop}%
\bibitem [{\citenamefont {Parsons}\ \emph {et~al.}(2016)\citenamefont
  {Parsons}, \citenamefont {Mazurenko}, \citenamefont {Chiu}, \citenamefont
  {Ji}, \citenamefont {Greif},\ and\ \citenamefont {Greiner}}]{Parsons2016}%
  \BibitemOpen
  \bibfield  {author} {\bibinfo {author} {\bibfnamefont {M.~F.}\ \bibnamefont
  {Parsons}}, \bibinfo {author} {\bibfnamefont {A.}~\bibnamefont {Mazurenko}},
  \bibinfo {author} {\bibfnamefont {C.~S.}\ \bibnamefont {Chiu}}, \bibinfo
  {author} {\bibfnamefont {G.}~\bibnamefont {Ji}}, \bibinfo {author}
  {\bibfnamefont {D.}~\bibnamefont {Greif}}, \ and\ \bibinfo {author}
  {\bibfnamefont {M.}~\bibnamefont {Greiner}},\ }\bibfield  {title} {\enquote
  {\bibinfo {title} {Site-resolved measurement of the spin-correlation function
  in the {Fermi-Hubbard} model},}\ }\href {\doibase 10.1126/science.aag1430}
  {\bibfield  {journal} {\bibinfo  {journal} {Science}\ }\textbf {\bibinfo
  {volume} {353}},\ \bibinfo {pages} {1253--1256} (\bibinfo {year}
  {2016})}\BibitemShut {NoStop}%
\bibitem [{\citenamefont {Cheuk}\ \emph
  {et~al.}(2016{\natexlab{b}})\citenamefont {Cheuk}, \citenamefont {Nichols},
  \citenamefont {Lawrence}, \citenamefont {Okan}, \citenamefont {Zhang},
  \citenamefont {Khatami}, \citenamefont {Trivedi}, \citenamefont {Paiva},
  \citenamefont {Rigol},\ and\ \citenamefont {Zwierlein}}]{Cheuk2016}%
  \BibitemOpen
  \bibfield  {author} {\bibinfo {author} {\bibfnamefont {L.~W.}\ \bibnamefont
  {Cheuk}}, \bibinfo {author} {\bibfnamefont {M.~A.}\ \bibnamefont {Nichols}},
  \bibinfo {author} {\bibfnamefont {K.~R.}\ \bibnamefont {Lawrence}}, \bibinfo
  {author} {\bibfnamefont {M.}~\bibnamefont {Okan}}, \bibinfo {author}
  {\bibfnamefont {H.}~\bibnamefont {Zhang}}, \bibinfo {author} {\bibfnamefont
  {E.}~\bibnamefont {Khatami}}, \bibinfo {author} {\bibfnamefont
  {N.}~\bibnamefont {Trivedi}}, \bibinfo {author} {\bibfnamefont
  {T.}~\bibnamefont {Paiva}}, \bibinfo {author} {\bibfnamefont
  {M.}~\bibnamefont {Rigol}}, \ and\ \bibinfo {author} {\bibfnamefont {M.~W.}\
  \bibnamefont {Zwierlein}},\ }\bibfield  {title} {\enquote {\bibinfo {title}
  {Observation of spatial charge and spin correlations in the {2D}
  {F}ermi-{H}ubbard model},}\ }\href {\doibase 10.1126/science.aag3349}
  {\bibfield  {journal} {\bibinfo  {journal} {Science}\ }\textbf {\bibinfo
  {volume} {353}},\ \bibinfo {pages} {1260--1264} (\bibinfo {year}
  {2016}{\natexlab{b}})}\BibitemShut {NoStop}%
\bibitem [{\citenamefont {Drewes}\ \emph {et~al.}(2016)\citenamefont {Drewes},
  \citenamefont {Cocchi}, \citenamefont {Miller}, \citenamefont {Chan},
  \citenamefont {Pertot}, \citenamefont {Brennecke},\ and\ \citenamefont
  {K\"ohl}}]{Drewes2016}%
  \BibitemOpen
  \bibfield  {author} {\bibinfo {author} {\bibfnamefont {J.~H.}\ \bibnamefont
  {Drewes}}, \bibinfo {author} {\bibfnamefont {E.}~\bibnamefont {Cocchi}},
  \bibinfo {author} {\bibfnamefont {L.~A.}\ \bibnamefont {Miller}}, \bibinfo
  {author} {\bibfnamefont {C.~F.}\ \bibnamefont {Chan}}, \bibinfo {author}
  {\bibfnamefont {D.}~\bibnamefont {Pertot}}, \bibinfo {author} {\bibfnamefont
  {F.}~\bibnamefont {Brennecke}}, \ and\ \bibinfo {author} {\bibfnamefont
  {M.}~\bibnamefont {K\"ohl}},\ }\bibfield  {title} {\enquote {\bibinfo {title}
  {Thermodynamics versus local density fluctuations in the
  metal--{M}ott-insulator crossover},}\ }\href {\doibase
  10.1103/PhysRevLett.117.135301} {\bibfield  {journal} {\bibinfo  {journal}
  {Phys. Rev. Lett.}\ }\textbf {\bibinfo {volume} {117}},\ \bibinfo {pages}
  {135301} (\bibinfo {year} {2016})}\BibitemShut {NoStop}%
\bibitem [{\citenamefont {Drewes}\ \emph {et~al.}(2017)\citenamefont {Drewes},
  \citenamefont {Miller}, \citenamefont {Cocchi}, \citenamefont {Chan},
  \citenamefont {Wurz}, \citenamefont {Gall}, \citenamefont {Pertot},
  \citenamefont {Brennecke},\ and\ \citenamefont {K\"ohl}}]{Drewes2017}%
  \BibitemOpen
  \bibfield  {author} {\bibinfo {author} {\bibfnamefont {J.~H.}\ \bibnamefont
  {Drewes}}, \bibinfo {author} {\bibfnamefont {L.~A.}\ \bibnamefont {Miller}},
  \bibinfo {author} {\bibfnamefont {E.}~\bibnamefont {Cocchi}}, \bibinfo
  {author} {\bibfnamefont {C.~F.}\ \bibnamefont {Chan}}, \bibinfo {author}
  {\bibfnamefont {N.}~\bibnamefont {Wurz}}, \bibinfo {author} {\bibfnamefont
  {M.}~\bibnamefont {Gall}}, \bibinfo {author} {\bibfnamefont {D.}~\bibnamefont
  {Pertot}}, \bibinfo {author} {\bibfnamefont {F.}~\bibnamefont {Brennecke}}, \
  and\ \bibinfo {author} {\bibfnamefont {M.}~\bibnamefont {K\"ohl}},\
  }\bibfield  {title} {\enquote {\bibinfo {title} {Antiferromagnetic
  correlations in two-dimensional fermionic {M}ott-insulating and metallic
  phases},}\ }\href {\doibase 10.1103/PhysRevLett.118.170401} {\bibfield
  {journal} {\bibinfo  {journal} {Phys. Rev. Lett.}\ }\textbf {\bibinfo
  {volume} {118}},\ \bibinfo {pages} {170401} (\bibinfo {year}
  {2017})}\BibitemShut {NoStop}%
\bibitem [{\citenamefont {Mazurenko}\ \emph {et~al.}(2017)\citenamefont
  {Mazurenko}, \citenamefont {Chiu}, \citenamefont {Ji}, \citenamefont
  {Parsons}, \citenamefont {Kan\'asz-Nagy}, \citenamefont {Schmidt},
  \citenamefont {Grusdt}, \citenamefont {Demler}, \citenamefont {Greif},\ and\
  \citenamefont {Greiner}}]{Mazurenko2016}%
  \BibitemOpen
  \bibfield  {author} {\bibinfo {author} {\bibfnamefont {A.}~\bibnamefont
  {Mazurenko}}, \bibinfo {author} {\bibfnamefont {C.~S.}\ \bibnamefont {Chiu}},
  \bibinfo {author} {\bibfnamefont {G.}~\bibnamefont {Ji}}, \bibinfo {author}
  {\bibfnamefont {M.~F.}\ \bibnamefont {Parsons}}, \bibinfo {author}
  {\bibfnamefont {M.}~\bibnamefont {Kan\'asz-Nagy}}, \bibinfo {author}
  {\bibfnamefont {R.}~\bibnamefont {Schmidt}}, \bibinfo {author} {\bibfnamefont
  {F.}~\bibnamefont {Grusdt}}, \bibinfo {author} {\bibfnamefont
  {E.}~\bibnamefont {Demler}}, \bibinfo {author} {\bibfnamefont
  {D.}~\bibnamefont {Greif}}, \ and\ \bibinfo {author} {\bibfnamefont
  {M.}~\bibnamefont {Greiner}},\ }\bibfield  {title} {\enquote {\bibinfo
  {title} {A cold-atom {Fermi-Hubbard} antiferromagnet},}\ }\href {\doibase
  10.1038/nature22362} {\bibfield  {journal} {\bibinfo  {journal} {Nature}\
  }\textbf {\bibinfo {volume} {545}},\ \bibinfo {pages} {462--466} (\bibinfo
  {year} {2017})}\BibitemShut {NoStop}%
\bibitem [{\citenamefont {Cocchi}\ \emph {et~al.}(2017)\citenamefont {Cocchi},
  \citenamefont {Miller}, \citenamefont {Drewes}, \citenamefont {Chan},
  \citenamefont {Pertot}, \citenamefont {Brennecke},\ and\ \citenamefont
  {K\"ohl}}]{Cocchi2017}%
  \BibitemOpen
  \bibfield  {author} {\bibinfo {author} {\bibfnamefont {E.}~\bibnamefont
  {Cocchi}}, \bibinfo {author} {\bibfnamefont {L.~A.}\ \bibnamefont {Miller}},
  \bibinfo {author} {\bibfnamefont {J.~H.}\ \bibnamefont {Drewes}}, \bibinfo
  {author} {\bibfnamefont {C.~F.}\ \bibnamefont {Chan}}, \bibinfo {author}
  {\bibfnamefont {D.}~\bibnamefont {Pertot}}, \bibinfo {author} {\bibfnamefont
  {F.}~\bibnamefont {Brennecke}}, \ and\ \bibinfo {author} {\bibfnamefont
  {M.}~\bibnamefont {K\"ohl}},\ }\bibfield  {title} {\enquote {\bibinfo {title}
  {Measuring entropy and short-range correlations in the two-dimensional
  {H}ubbard model},}\ }\href {\doibase 10.1103/PhysRevX.7.031025} {\bibfield
  {journal} {\bibinfo  {journal} {Phys. Rev. X}\ }\textbf {\bibinfo {volume}
  {7}},\ \bibinfo {pages} {031025} (\bibinfo {year} {2017})}\BibitemShut
  {NoStop}%
\bibitem [{\citenamefont {Brown}\ \emph {et~al.}(2017)\citenamefont {Brown},
  \citenamefont {Mitra}, \citenamefont {Guardado-Sanchez}, \citenamefont
  {Schau{\ss}}, \citenamefont {Kondov}, \citenamefont {Khatami}, \citenamefont
  {Paiva}, \citenamefont {Trivedi}, \citenamefont {Huse},\ and\ \citenamefont
  {Bakr}}]{Brown2017}%
  \BibitemOpen
  \bibfield  {author} {\bibinfo {author} {\bibfnamefont {P.~T.}\ \bibnamefont
  {Brown}}, \bibinfo {author} {\bibfnamefont {D.}~\bibnamefont {Mitra}},
  \bibinfo {author} {\bibfnamefont {E.}~\bibnamefont {Guardado-Sanchez}},
  \bibinfo {author} {\bibfnamefont {P.}~\bibnamefont {Schau{\ss}}}, \bibinfo
  {author} {\bibfnamefont {S.~S.}\ \bibnamefont {Kondov}}, \bibinfo {author}
  {\bibfnamefont {E.}~\bibnamefont {Khatami}}, \bibinfo {author} {\bibfnamefont
  {T.}~\bibnamefont {Paiva}}, \bibinfo {author} {\bibfnamefont
  {N.}~\bibnamefont {Trivedi}}, \bibinfo {author} {\bibfnamefont {D.~A.}\
  \bibnamefont {Huse}}, \ and\ \bibinfo {author} {\bibfnamefont {W.~S.}\
  \bibnamefont {Bakr}},\ }\bibfield  {title} {\enquote {\bibinfo {title}
  {Spin-imbalance in a {2D Fermi-Hubbard} system},}\ }\href {\doibase
  10.1126/science.aam7838} {\bibfield  {journal} {\bibinfo  {journal}
  {Science}\ }\textbf {\bibinfo {volume} {357}},\ \bibinfo {pages} {1385--1388}
  (\bibinfo {year} {2017})}\BibitemShut {NoStop}%
\bibitem [{\citenamefont {Rigol}\ \emph {et~al.}(2003)\citenamefont {Rigol},
  \citenamefont {Muramatsu}, \citenamefont {Batrouni},\ and\ \citenamefont
  {Scalettar}}]{rigol_muramatsu_03}%
  \BibitemOpen
  \bibfield  {author} {\bibinfo {author} {\bibfnamefont {M.}~\bibnamefont
  {Rigol}}, \bibinfo {author} {\bibfnamefont {A.}~\bibnamefont {Muramatsu}},
  \bibinfo {author} {\bibfnamefont {G.~G.}\ \bibnamefont {Batrouni}}, \ and\
  \bibinfo {author} {\bibfnamefont {R.~T.}\ \bibnamefont {Scalettar}},\
  }\bibfield  {title} {\enquote {\bibinfo {title} {Local quantum criticality in
  confined fermions on optical lattices},}\ }\href {\doibase
  10.1103/PhysRevLett.91.130403} {\bibfield  {journal} {\bibinfo  {journal}
  {Phys. Rev. Lett.}\ }\textbf {\bibinfo {volume} {91}},\ \bibinfo {pages}
  {130403} (\bibinfo {year} {2003})}\BibitemShut {NoStop}%
\bibitem [{\citenamefont {Rigol}\ and\ \citenamefont
  {Muramatsu}(2004)}]{rigol_muramatsu_04a}%
  \BibitemOpen
  \bibfield  {author} {\bibinfo {author} {\bibfnamefont {M.}~\bibnamefont
  {Rigol}}\ and\ \bibinfo {author} {\bibfnamefont {A.}~\bibnamefont
  {Muramatsu}},\ }\bibfield  {title} {\enquote {\bibinfo {title} {Quantum
  {M}onte {C}arlo study of confined fermions in one-dimensional optical
  lattices},}\ }\href {\doibase 10.1103/PhysRevA.69.053612} {\bibfield
  {journal} {\bibinfo  {journal} {Phys. Rev. A}\ }\textbf {\bibinfo {volume}
  {69}},\ \bibinfo {pages} {053612} (\bibinfo {year} {2004})}\BibitemShut
  {NoStop}%
\bibitem [{\citenamefont {Chiesa}\ \emph {et~al.}(2011)\citenamefont {Chiesa},
  \citenamefont {Varney}, \citenamefont {Rigol},\ and\ \citenamefont
  {Scalettar}}]{chiesa_varney_11}%
  \BibitemOpen
  \bibfield  {author} {\bibinfo {author} {\bibfnamefont {S.}~\bibnamefont
  {Chiesa}}, \bibinfo {author} {\bibfnamefont {C.~N.}\ \bibnamefont {Varney}},
  \bibinfo {author} {\bibfnamefont {M.}~\bibnamefont {Rigol}}, \ and\ \bibinfo
  {author} {\bibfnamefont {R.~T.}\ \bibnamefont {Scalettar}},\ }\bibfield
  {title} {\enquote {\bibinfo {title} {Magnetism and pairing of two-dimensional
  trapped fermions},}\ }\href {\doibase 10.1103/PhysRevLett.106.035301}
  {\bibfield  {journal} {\bibinfo  {journal} {Phys. Rev. Lett.}\ }\textbf
  {\bibinfo {volume} {106}},\ \bibinfo {pages} {035301} (\bibinfo {year}
  {2011})}\BibitemShut {NoStop}%
\bibitem [{\citenamefont {Machida}\ \emph {et~al.}(2004)\citenamefont
  {Machida}, \citenamefont {Yamada}, \citenamefont {Ohashi},\ and\
  \citenamefont {Matsumoto}}]{Machida2004}%
  \BibitemOpen
  \bibfield  {author} {\bibinfo {author} {\bibfnamefont {M.}~\bibnamefont
  {Machida}}, \bibinfo {author} {\bibfnamefont {S.}~\bibnamefont {Yamada}},
  \bibinfo {author} {\bibfnamefont {Y.}~\bibnamefont {Ohashi}}, \ and\ \bibinfo
  {author} {\bibfnamefont {H.}~\bibnamefont {Matsumoto}},\ }\bibfield  {title}
  {\enquote {\bibinfo {title} {Novel superfluidity in a trapped gas of {F}ermi
  atoms with repulsive interaction loaded on an optical lattice},}\ }\href
  {\doibase 10.1103/PhysRevLett.93.200402} {\bibfield  {journal} {\bibinfo
  {journal} {Phys. Rev. Lett.}\ }\textbf {\bibinfo {volume} {93}},\ \bibinfo
  {pages} {200402} (\bibinfo {year} {2004})}\BibitemShut {NoStop}%
\bibitem [{\citenamefont {Rigol}\ \emph {et~al.}(2005)\citenamefont {Rigol},
  \citenamefont {Manmana}, \citenamefont {Muramatsu}, \citenamefont
  {Scalettar}, \citenamefont {Singh},\ and\ \citenamefont
  {Wessel}}]{Rigol2005}%
  \BibitemOpen
  \bibfield  {author} {\bibinfo {author} {\bibfnamefont {M.}~\bibnamefont
  {Rigol}}, \bibinfo {author} {\bibfnamefont {S.~R.}\ \bibnamefont {Manmana}},
  \bibinfo {author} {\bibfnamefont {A.}~\bibnamefont {Muramatsu}}, \bibinfo
  {author} {\bibfnamefont {R.~T.}\ \bibnamefont {Scalettar}}, \bibinfo {author}
  {\bibfnamefont {R.~R.~P.}\ \bibnamefont {Singh}}, \ and\ \bibinfo {author}
  {\bibfnamefont {S.}~\bibnamefont {Wessel}},\ }\bibfield  {title} {\enquote
  {\bibinfo {title} {Comment on ``{N}ovel superfluidity in a trapped gas of
  {F}ermi atoms with repulsive interaction loaded on an optical lattice''},}\
  }\href {\doibase 10.1103/PhysRevLett.95.218901} {\bibfield  {journal}
  {\bibinfo  {journal} {Phys. Rev. Lett.}\ }\textbf {\bibinfo {volume} {95}},\
  \bibinfo {pages} {218901} (\bibinfo {year} {2005})}\BibitemShut {NoStop}%
\bibitem [{\citenamefont {Machida}\ \emph {et~al.}(2005)\citenamefont
  {Machida}, \citenamefont {Yamada}, \citenamefont {Ohashi},\ and\
  \citenamefont {Matsumoto}}]{Machida2005}%
  \BibitemOpen
  \bibfield  {author} {\bibinfo {author} {\bibfnamefont {M.}~\bibnamefont
  {Machida}}, \bibinfo {author} {\bibfnamefont {S.}~\bibnamefont {Yamada}},
  \bibinfo {author} {\bibfnamefont {Y.}~\bibnamefont {Ohashi}}, \ and\ \bibinfo
  {author} {\bibfnamefont {H.}~\bibnamefont {Matsumoto}},\ }\bibfield  {title}
  {\enquote {\bibinfo {title} {Machida {\it et al.}~reply:},}\ }\href {\doibase
  10.1103/PhysRevLett.95.218902} {\bibfield  {journal} {\bibinfo  {journal}
  {Phys. Rev. Lett.}\ }\textbf {\bibinfo {volume} {95}},\ \bibinfo {pages}
  {218902} (\bibinfo {year} {2005})}\BibitemShut {NoStop}%
\bibitem [{\citenamefont {Emery}\ \emph {et~al.}(1997)\citenamefont {Emery},
  \citenamefont {Kivelson},\ and\ \citenamefont {Zachar}}]{Emery1997}%
  \BibitemOpen
  \bibfield  {author} {\bibinfo {author} {\bibfnamefont {V.~J.}\ \bibnamefont
  {Emery}}, \bibinfo {author} {\bibfnamefont {S.~A.}\ \bibnamefont {Kivelson}},
  \ and\ \bibinfo {author} {\bibfnamefont {O.}~\bibnamefont {Zachar}},\
  }\bibfield  {title} {\enquote {\bibinfo {title} {Spin-gap proximity effect
  mechanism of high-temperature superconductivity},}\ }\href {\doibase
  10.1103/PhysRevB.56.6120} {\bibfield  {journal} {\bibinfo  {journal} {Phys.
  Rev. B}\ }\textbf {\bibinfo {volume} {56}},\ \bibinfo {pages} {6120--6147}
  (\bibinfo {year} {1997})}\BibitemShut {NoStop}%
\bibitem [{\citenamefont {Tsvelik}(2016{\natexlab{a}})}]{tsvelik2016a}%
  \BibitemOpen
  \bibfield  {author} {\bibinfo {author} {\bibfnamefont {A.~M.}\ \bibnamefont
  {Tsvelik}},\ }\bibfield  {title} {\enquote {\bibinfo {title} {Universality
  classes of order parameters composed of many-body bound states},}\ }\href
  {\doibase 10.1103/PhysRevB.94.205141} {\bibfield  {journal} {\bibinfo
  {journal} {Phys. Rev. B}\ }\textbf {\bibinfo {volume} {94}},\ \bibinfo
  {pages} {205141} (\bibinfo {year} {2016}{\natexlab{a}})}\BibitemShut
  {NoStop}%
\bibitem [{\citenamefont {Tsvelik}(2016{\natexlab{b}})}]{tsvelik2016b}%
  \BibitemOpen
  \bibfield  {author} {\bibinfo {author} {\bibfnamefont {A.~M.}\ \bibnamefont
  {Tsvelik}},\ }\bibfield  {title} {\enquote {\bibinfo {title} {Fractionalized
  {F}ermi liquid in a {Kondo-Heisenberg} model},}\ }\href {\doibase
  10.1103/PhysRevB.94.165114} {\bibfield  {journal} {\bibinfo  {journal} {Phys.
  Rev. B}\ }\textbf {\bibinfo {volume} {94}},\ \bibinfo {pages} {165114}
  (\bibinfo {year} {2016}{\natexlab{b}})}\BibitemShut {NoStop}%
\bibitem [{\citenamefont {Hayward}\ \emph {et~al.}(1995)\citenamefont
  {Hayward}, \citenamefont {Poilblanc}, \citenamefont {Noack}, \citenamefont
  {Scalapino},\ and\ \citenamefont {Hanke}}]{Hayward1995}%
  \BibitemOpen
  \bibfield  {author} {\bibinfo {author} {\bibfnamefont {C.~A.}\ \bibnamefont
  {Hayward}}, \bibinfo {author} {\bibfnamefont {D.}~\bibnamefont {Poilblanc}},
  \bibinfo {author} {\bibfnamefont {R.~M.}\ \bibnamefont {Noack}}, \bibinfo
  {author} {\bibfnamefont {D.~J.}\ \bibnamefont {Scalapino}}, \ and\ \bibinfo
  {author} {\bibfnamefont {W.}~\bibnamefont {Hanke}},\ }\bibfield  {title}
  {\enquote {\bibinfo {title} {Evidence for a superfluid density in
  {$\mathit{t}$-$\mathit{J}$} ladders},}\ }\href {\doibase
  10.1103/PhysRevLett.75.926} {\bibfield  {journal} {\bibinfo  {journal} {Phys.
  Rev. Lett.}\ }\textbf {\bibinfo {volume} {75}},\ \bibinfo {pages} {926--929}
  (\bibinfo {year} {1995})}\BibitemShut {NoStop}%
\bibitem [{\citenamefont {Poilblanc}\ \emph {et~al.}(2003)\citenamefont
  {Poilblanc}, \citenamefont {Scalapino},\ and\ \citenamefont
  {Capponi}}]{Poilblanc2003}%
  \BibitemOpen
  \bibfield  {author} {\bibinfo {author} {\bibfnamefont {D.}~\bibnamefont
  {Poilblanc}}, \bibinfo {author} {\bibfnamefont {D.~J.}\ \bibnamefont
  {Scalapino}}, \ and\ \bibinfo {author} {\bibfnamefont {S.}~\bibnamefont
  {Capponi}},\ }\bibfield  {title} {\enquote {\bibinfo {title} {Superconducting
  gap for a two-leg {$t\mathrm{\text{\ensuremath{-}}}J$} ladder},}\ }\href
  {\doibase 10.1103/PhysRevLett.91.137203} {\bibfield  {journal} {\bibinfo
  {journal} {Phys. Rev. Lett.}\ }\textbf {\bibinfo {volume} {91}},\ \bibinfo
  {pages} {137203} (\bibinfo {year} {2003})}\BibitemShut {NoStop}%
\bibitem [{\citenamefont {White}\ and\ \citenamefont
  {Scalapino}(1997)}]{White1997}%
  \BibitemOpen
  \bibfield  {author} {\bibinfo {author} {\bibfnamefont {S.~R.}\ \bibnamefont
  {White}}\ and\ \bibinfo {author} {\bibfnamefont {D.~J.}\ \bibnamefont
  {Scalapino}},\ }\bibfield  {title} {\enquote {\bibinfo {title} {Ground states
  of the doped four-leg $t$-${J}$ ladder},}\ }\href {\doibase
  10.1103/PhysRevB.55.R14701} {\bibfield  {journal} {\bibinfo  {journal} {Phys.
  Rev. B}\ }\textbf {\bibinfo {volume} {55}},\ \bibinfo {pages}
  {R14701--R14704} (\bibinfo {year} {1997})}\BibitemShut {NoStop}%
\bibitem [{\citenamefont {White}\ and\ \citenamefont
  {Scalapino}(1998)}]{White1998}%
  \BibitemOpen
  \bibfield  {author} {\bibinfo {author} {\bibfnamefont {S.~R.}\ \bibnamefont
  {White}}\ and\ \bibinfo {author} {\bibfnamefont {D.~J.}\ \bibnamefont
  {Scalapino}},\ }\bibfield  {title} {\enquote {\bibinfo {title} {Density
  matrix renormalization group study of the striped phase in the {2D}
  $\mathit{t}\ensuremath{-}\mathit{J}$ model},}\ }\href {\doibase
  10.1103/PhysRevLett.80.1272} {\bibfield  {journal} {\bibinfo  {journal}
  {Phys. Rev. Lett.}\ }\textbf {\bibinfo {volume} {80}},\ \bibinfo {pages}
  {1272--1275} (\bibinfo {year} {1998})}\BibitemShut {NoStop}%
\bibitem [{\citenamefont {White}\ and\ \citenamefont
  {Scalapino}(2003)}]{White2003}%
  \BibitemOpen
  \bibfield  {author} {\bibinfo {author} {\bibfnamefont {S.~R.}\ \bibnamefont
  {White}}\ and\ \bibinfo {author} {\bibfnamefont {D.~J.}\ \bibnamefont
  {Scalapino}},\ }\bibfield  {title} {\enquote {\bibinfo {title} {Stripes on a
  6-leg {H}ubbard ladder},}\ }\href {\doibase 10.1103/PhysRevLett.91.136403}
  {\bibfield  {journal} {\bibinfo  {journal} {Phys. Rev. Lett.}\ }\textbf
  {\bibinfo {volume} {91}},\ \bibinfo {pages} {136403} (\bibinfo {year}
  {2003})}\BibitemShut {NoStop}%
\bibitem [{\citenamefont {Chang}\ and\ \citenamefont
  {Affleck}(2007)}]{Chang2007}%
  \BibitemOpen
  \bibfield  {author} {\bibinfo {author} {\bibfnamefont {M.-S.}\ \bibnamefont
  {Chang}}\ and\ \bibinfo {author} {\bibfnamefont {I.}~\bibnamefont
  {Affleck}},\ }\bibfield  {title} {\enquote {\bibinfo {title} {Bipairing and
  the stripe phase in four-leg {H}ubbard ladders},}\ }\href {\doibase
  10.1103/PhysRevB.76.054521} {\bibfield  {journal} {\bibinfo  {journal} {Phys.
  Rev. B}\ }\textbf {\bibinfo {volume} {76}},\ \bibinfo {pages} {054521}
  (\bibinfo {year} {2007})}\BibitemShut {NoStop}%
\bibitem [{\citenamefont {Rigol}\ \emph {et~al.}(2007)\citenamefont {Rigol},
  \citenamefont {Bryant},\ and\ \citenamefont {Singh}}]{rigol_bryant_07b}%
  \BibitemOpen
  \bibfield  {author} {\bibinfo {author} {\bibfnamefont {M.}~\bibnamefont
  {Rigol}}, \bibinfo {author} {\bibfnamefont {T.}~\bibnamefont {Bryant}}, \
  and\ \bibinfo {author} {\bibfnamefont {R.~R.~P.}\ \bibnamefont {Singh}},\
  }\bibfield  {title} {\enquote {\bibinfo {title} {Numerical linked-cluster
  algorithms. {II. $t\text{-}J$} models on the square lattice},}\ }\href
  {\doibase 10.1103/PhysRevE.75.061119} {\bibfield  {journal} {\bibinfo
  {journal} {Phys. Rev. E}\ }\textbf {\bibinfo {volume} {75}},\ \bibinfo
  {pages} {061119} (\bibinfo {year} {2007})}\BibitemShut {NoStop}%
\bibitem [{\citenamefont {Rigol}\ \emph {et~al.}(2009)\citenamefont {Rigol},
  \citenamefont {Shastry},\ and\ \citenamefont {Haas}}]{rigol_shastry_09b}%
  \BibitemOpen
  \bibfield  {author} {\bibinfo {author} {\bibfnamefont {M.}~\bibnamefont
  {Rigol}}, \bibinfo {author} {\bibfnamefont {B.~S.}\ \bibnamefont {Shastry}},
  \ and\ \bibinfo {author} {\bibfnamefont {S.}~\bibnamefont {Haas}},\
  }\bibfield  {title} {\enquote {\bibinfo {title} {Fidelity and
  superconductivity in two-dimensional {$t\text{-}J$} models},}\ }\href
  {\doibase 10.1103/PhysRevB.80.094529} {\bibfield  {journal} {\bibinfo
  {journal} {Phys. Rev. B}\ }\textbf {\bibinfo {volume} {80}},\ \bibinfo
  {pages} {094529} (\bibinfo {year} {2009})}\BibitemShut {NoStop}%
\bibitem [{\citenamefont {Moreno}\ \emph {et~al.}(2011)\citenamefont {Moreno},
  \citenamefont {Muramatsu},\ and\ \citenamefont {Manmana}}]{Moreno2011}%
  \BibitemOpen
  \bibfield  {author} {\bibinfo {author} {\bibfnamefont {A.}~\bibnamefont
  {Moreno}}, \bibinfo {author} {\bibfnamefont {A.}~\bibnamefont {Muramatsu}}, \
  and\ \bibinfo {author} {\bibfnamefont {S.~R.}\ \bibnamefont {Manmana}},\
  }\bibfield  {title} {\enquote {\bibinfo {title} {Ground-state phase diagram
  of the one-dimensional {$t$-$J$} model},}\ }\href {\doibase
  10.1103/PhysRevB.83.205113} {\bibfield  {journal} {\bibinfo  {journal} {Phys.
  Rev. B}\ }\textbf {\bibinfo {volume} {83}},\ \bibinfo {pages} {205113}
  (\bibinfo {year} {2011})}\BibitemShut {NoStop}%
\bibitem [{\citenamefont {Corboz}\ \emph {et~al.}(2011)\citenamefont {Corboz},
  \citenamefont {White}, \citenamefont {Vidal},\ and\ \citenamefont
  {Troyer}}]{Corboz2011}%
  \BibitemOpen
  \bibfield  {author} {\bibinfo {author} {\bibfnamefont {P.}~\bibnamefont
  {Corboz}}, \bibinfo {author} {\bibfnamefont {S.~R.}\ \bibnamefont {White}},
  \bibinfo {author} {\bibfnamefont {G.}~\bibnamefont {Vidal}}, \ and\ \bibinfo
  {author} {\bibfnamefont {M.}~\bibnamefont {Troyer}},\ }\bibfield  {title}
  {\enquote {\bibinfo {title} {Stripes in the two-dimensional $t$-${J}$ model
  with infinite projected entangled-pair states},}\ }\href {\doibase
  10.1103/PhysRevB.84.041108} {\bibfield  {journal} {\bibinfo  {journal} {Phys.
  Rev. B}\ }\textbf {\bibinfo {volume} {84}},\ \bibinfo {pages} {041108}
  (\bibinfo {year} {2011})}\BibitemShut {NoStop}%
\bibitem [{\citenamefont {Gorshkov}\ \emph
  {et~al.}(2011{\natexlab{a}})\citenamefont {Gorshkov}, \citenamefont
  {Manmana}, \citenamefont {Chen}, \citenamefont {Ye}, \citenamefont {Demler},
  \citenamefont {Lukin},\ and\ \citenamefont {Rey}}]{Gorshkov2011}%
  \BibitemOpen
  \bibfield  {author} {\bibinfo {author} {\bibfnamefont {A.~V.}\ \bibnamefont
  {Gorshkov}}, \bibinfo {author} {\bibfnamefont {S.~R.}\ \bibnamefont
  {Manmana}}, \bibinfo {author} {\bibfnamefont {G.}~\bibnamefont {Chen}},
  \bibinfo {author} {\bibfnamefont {J.}~\bibnamefont {Ye}}, \bibinfo {author}
  {\bibfnamefont {E.}~\bibnamefont {Demler}}, \bibinfo {author} {\bibfnamefont
  {M.~D.}\ \bibnamefont {Lukin}}, \ and\ \bibinfo {author} {\bibfnamefont
  {A.~M.}\ \bibnamefont {Rey}},\ }\bibfield  {title} {\enquote {\bibinfo
  {title} {Tunable superfluidity and quantum magnetism with ultracold polar
  molecules},}\ }\href {\doibase 10.1103/PhysRevLett.107.115301} {\bibfield
  {journal} {\bibinfo  {journal} {Phys. Rev. Lett.}\ }\textbf {\bibinfo
  {volume} {107}},\ \bibinfo {pages} {115301} (\bibinfo {year}
  {2011}{\natexlab{a}})}\BibitemShut {NoStop}%
\bibitem [{\citenamefont {Gorshkov}\ \emph
  {et~al.}(2011{\natexlab{b}})\citenamefont {Gorshkov}, \citenamefont
  {Manmana}, \citenamefont {Chen}, \citenamefont {Demler}, \citenamefont
  {Lukin},\ and\ \citenamefont {Rey}}]{Gorshkov2011a}%
  \BibitemOpen
  \bibfield  {author} {\bibinfo {author} {\bibfnamefont {A.~V.}\ \bibnamefont
  {Gorshkov}}, \bibinfo {author} {\bibfnamefont {S.~R.}\ \bibnamefont
  {Manmana}}, \bibinfo {author} {\bibfnamefont {G.}~\bibnamefont {Chen}},
  \bibinfo {author} {\bibfnamefont {E.}~\bibnamefont {Demler}}, \bibinfo
  {author} {\bibfnamefont {M.~D.}\ \bibnamefont {Lukin}}, \ and\ \bibinfo
  {author} {\bibfnamefont {A.~M.}\ \bibnamefont {Rey}},\ }\bibfield  {title}
  {\enquote {\bibinfo {title} {Quantum magnetism with polar alkali-metal
  dimers},}\ }\href {\doibase 10.1103/PhysRevA.84.033619} {\bibfield  {journal}
  {\bibinfo  {journal} {Phys. Rev. A}\ }\textbf {\bibinfo {volume} {84}},\
  \bibinfo {pages} {033619} (\bibinfo {year} {2011}{\natexlab{b}})}\BibitemShut
  {NoStop}%
\bibitem [{\citenamefont {White}(1992)}]{White1992}%
  \BibitemOpen
  \bibfield  {author} {\bibinfo {author} {\bibfnamefont {S.~R.}\ \bibnamefont
  {White}},\ }\bibfield  {title} {\enquote {\bibinfo {title} {Density matrix
  formulation for quantum renormalization groups},}\ }\href {\doibase
  10.1103/PhysRevLett.69.2863} {\bibfield  {journal} {\bibinfo  {journal}
  {Phys. Rev. Lett.}\ }\textbf {\bibinfo {volume} {69}},\ \bibinfo {pages}
  {2863} (\bibinfo {year} {1992})}\BibitemShut {NoStop}%
\bibitem [{\citenamefont {White}(1993)}]{White1993}%
  \BibitemOpen
  \bibfield  {author} {\bibinfo {author} {\bibfnamefont {S.~R.}\ \bibnamefont
  {White}},\ }\bibfield  {title} {\enquote {\bibinfo {title} {Density-matrix
  algorithms for quantum renormalization groups},}\ }\href {\doibase
  10.1103/PhysRevB.48.10345} {\bibfield  {journal} {\bibinfo  {journal} {Phys.
  Rev. B}\ }\textbf {\bibinfo {volume} {48}},\ \bibinfo {pages} {10345}
  (\bibinfo {year} {1993})}\BibitemShut {NoStop}%
\bibitem [{\citenamefont {Legeza}\ \emph {et~al.}(2003)\citenamefont {Legeza},
  \citenamefont {R\"oder},\ and\ \citenamefont {Hess}}]{Legeza2003}%
  \BibitemOpen
  \bibfield  {author} {\bibinfo {author} {\bibfnamefont {\"O.}\ \bibnamefont
  {Legeza}}, \bibinfo {author} {\bibfnamefont {J.}~\bibnamefont {R\"oder}}, \
  and\ \bibinfo {author} {\bibfnamefont {B.~A.}\ \bibnamefont {Hess}},\
  }\bibfield  {title} {\enquote {\bibinfo {title} {Controlling the accuracy of
  the density-matrix renormalization-group method: The dynamical block state
  selection approach},}\ }\href {\doibase 10.1103/PhysRevB.67.125114}
  {\bibfield  {journal} {\bibinfo  {journal} {Phys. Rev. B}\ }\textbf {\bibinfo
  {volume} {67}},\ \bibinfo {pages} {125114} (\bibinfo {year}
  {2003})}\BibitemShut {NoStop}%
\bibitem [{\citenamefont {Giamarchi}()}]{Giamarchibook}%
  \BibitemOpen
  \bibfield  {author} {\bibinfo {author} {\bibfnamefont {T.}~\bibnamefont
  {Giamarchi}},\ }\bibfield  {title} {\enquote {\bibinfo {title} {Quantum
  physics in one dimension},}\ }\href@noop {} {\bibinfo  {journal} {(Oxford
  University Press, Oxford, 2004)}\ }\BibitemShut {NoStop}%
\bibitem [{\citenamefont {Rohling}\ \emph {et~al.}(2018)\citenamefont
  {Rohling}, \citenamefont {Fj\ae{}rbu},\ and\ \citenamefont
  {Brataas}}]{Rohling2018}%
  \BibitemOpen
\bibfield  {journal} {  }\bibfield  {author} {\bibinfo {author} {\bibfnamefont
  {N.}~\bibnamefont {Rohling}}, \bibinfo {author} {\bibfnamefont {E.~L.}\
  \bibnamefont {Fj\ae{}rbu}}, \ and\ \bibinfo {author} {\bibfnamefont
  {A.}~\bibnamefont {Brataas}},\ }\bibfield  {title} {\enquote {\bibinfo
  {title} {Superconductivity induced by interfacial coupling to magnons},}\
  }\href {\doibase 10.1103/PhysRevB.97.115401} {\bibfield  {journal} {\bibinfo
  {journal} {Phys. Rev. B}\ }\textbf {\bibinfo {volume} {97}},\ \bibinfo
  {pages} {115401} (\bibinfo {year} {2018})}\BibitemShut {NoStop}%
\bibitem [{Note1()}]{Note1}%
  \BibitemOpen
  \bibinfo {note} {In ladders with an odd number of legs, we find that there is
  no novel interleg superconductivity of the type discussed here for ladders
  with an even number of legs. interleg singlet pairing between fermions in the
  outer legs is incompatible with antiferromagnetic correlations in an inner
  Mott insulating stripe with an odd number of legs, and triplet pairing does
  not appear to be energetically favorable in those
  configurations.}\BibitemShut {Stop}%
\end{thebibliography}%

\end{document}